%% file: main.tex
\documentclass[notitlepage,floatfix,aps,pra,reprint,superscriptaddress]{revtex4-2}
\usepackage{graphicx}
\usepackage{amsfonts,amssymb,amsthm,amsmath,amscd}
\usepackage{enumerate}
\usepackage{tikz}
\usepackage{xcolor}
\usepackage{algorithm}
\usepackage{algpseudocode}
\usepackage{braket}

\usepackage[colorlinks=true, citecolor=blue]{hyperref}
\usepackage{cleveref}

\usepackage{multirow}
\usepackage{array}
\usepackage{booktabs}
\newcommand{\EPFLQSE}{Centre for Quantum Science and Engineering, \'{E}cole Polytechnique F\'{e}d\'{e}rale de Lausanne (EPFL), Lausanne, Switzerland}
\newcommand{\EPFL}{Institute of Physics, \'{E}cole Polytechnique F\'{e}d\'{e}rale de
Lausanne (EPFL), Lausanne, Switzerland}
\newcommand{\KEIO}{Quantum Computing Center, Keio University, Hiyoshi 3-14-1, Kohoku-ku, Yokohama 223-8522, Japan}

\begin{document}
\title{Quantum-compatible Riemannian conjugate gradient for ground-state preparation}
\author{Andres Arcia-Lopez}
\author{Marek Gluza}
\affiliation{School of Physical and Mathematical Sciences, Nanyang Technological University, 637371, Singapore}
\author{Yudai Suzuki}
\affiliation{\EPFL}
\affiliation{\EPFLQSE}
\affiliation{\KEIO}
\author{Cyrus Mostajeran}
\affiliation{School of Physical and Mathematical Sciences, Nanyang Technological University, 637371, Singapore}

%\date{}

\begin{abstract}
Riemannian optimization provides a natural geometric framework for designing quantum algorithms for ground-state preparation, most notably through quantum imaginary-time evolution. However, existing geometry-aware approaches primarily employ steepest-descent updates, leaving more advanced methods from classical Riemannian optimization largely unexplored. In this work, we introduce quantum conjugate gradient descent (QCGD), a quantum-compatible implementation of Riemannian conjugate gradient that combines the steepest descent direction with information retained from previous search directions. In our numerical experiments, the proposed approach achieves faster convergence per iteration than the baseline double-bracket quantum imaginary-time evolution (DB-QITE) algorithm, and we quantify the associated circuit cost with gate counts produced by Qrisp. This improvement, however, comes at the cost of increased circuit depth for each update step, revealing a trade-off between convergence rate and implementation complexity. Nevertheless, since the circuit depth for those algorithms grows exponentially with the number of iterations, the proposed method may offer practical advantages, particularly in the early fault-tolerant regime.
\end{abstract}

\maketitle
\section{Introduction}
\label{introduction}

\emph{Ground-state preparation} (GSP) of Hamiltonians 
is a versatile and pivotal approach to studying the properties of chemicals \cite{mcardle2020quantum} and materials  \cite{alexeev2024quantum}. 
Similarly, this framework also proves useful in tackling problems in combinatorial optimization \cite{pelofskeShortDepthQAOACircuits2023,sciorilliLargescaleQuantumOptimization2025}, machine learning \cite{dunjkoNonreviewQuantumMachine2020,choMachineLearningQuantum2024,panDeepQuantumNeural2023,chenEmpoweringDeepNeural2024}, and certain computer science tasks such as Max-Cut, graph-partitioning, scheduling, and resource allocation \cite{lucasIsingFormulationsMany2014,farhi2014quantum,mohseniIsingMachinesHardware2022,jiangEfficientCombinatorialOptimization2023,bybeeEfficientOptimizationHigherorder2023,zhangReviewIsingMachines2024a,siEnergyefficientSuperparamagneticIsing2024,ushijima-mwesigwaGraphPartitioningUsing2017,pelofskeDecompositionAlgorithmsSolving2021,bozejkoOptimalSolvingScheduling2025,stollenwerkQuantumAnnealingApplied2020,fuSolvingFlexibleJobShop2025,ajagekarHybridClassicalQuantumOptimization2022,ohyamaResourceAllocationOptimization2023,buttQuantumInspiredResourceOptimization2025}.  However, for a general Hamiltonian, GSP is not only NP-hard \cite{barahona1982computational}, but also QMA-complete \cite{GottesmanQMA,Kempe-SIAM-2006,Bausch,Kempe-QIC-2003}, so it is a challenging task for both classical and quantum devices. Nevertheless, quantum computation is expected to offer significant advantages over classical approaches for important problem classes \cite{cade2020strategies,wu2024variational,AlexiaEnergyInitiative}. In fact, a quantum approach could not only provide practical \cite{NatafMilaSUN} improvements but also overcome the conceptual \cite{Feynman} and fundamental \cite{ETH_SAT,strassen1969gaussian} challenges of classically implemented GSP.

Thus far, there exist various classes of algorithms to execute GSP on both near-term \cite{avqite,mcArdle2019,nishi2021,yeter2020practical,motta2020determining,yeter2021benchmarking,Gomes2020,Huang2023,MottaAdiabatic,LinQPE,larose2019variational,Zeng_2021,parrish2019quantum,stair2020multireference,kokail2019self,google2020hartree,wiersemaOptimizingQuantumCircuits2023b,robbiatiDBQA2026,KishorRevModPhys.94.015004,cerezo2021variational,PreparationMPS2,PreparationMPS} and fault-tolerant \cite{nearoptimalground_Lin_2020,ge2019faster,gilyen2019quantum,fragmented_QITE2024,dong2022ground,epperly2021theory,temme2011quantum,kitaev1995quantum,brassard2002quantum,tan2020quantum,Low2019hamiltonian,PhysRevA.90.022305,gluza2024doublebracket,PoulinWocjan,motlagh2024ground,Kirby2023exactefficient,ding2024single,chen2025efficient,chen2024local} quantum hardware. One such class stems from the open problem of designing efficient quantum circuits that implement \emph{imaginary-time evolution} (ITE), Eq.~\eqref{introduction.eq.ITE}. These approaches range from hybrid quantum-classical optimization schemes \cite{motta2020determining,avqite,mcArdle2019,nishi2021,yeter2020practical,yeter2021benchmarking,Gomes2020} to block-encoding methods~\cite{fragmented_QITE2024}. However, important scalability \cite{larocca2024review, cerezo2023does} and hardware \cite{fragmented_QITE2024} issues plague these algorithms and motivate the pursuit of better, or entirely different, ideas. Consequently, designing efficient and scalable GSP quantum algorithms remains an open problem.

A novel perspective for motivating quantum algorithm design is that GSP strategies can naturally be formulated as an optimization problem that minimises the energy expectation value over a curved space. Because quantum states are constrained by normalisation and phase redundancy, it helps \cite{pervezRiemannianGradientDescentbased2025} to incorporate this geometry into the optimization formulation. For instance, recent work \cite{gluza_DB_QITE_2024} proposes \emph{double bracket quantum imaginary-time evolution}  (DB-QITE) as an efficient method better suited to current quantum hardware, see also~\cite{mcmahon2026equating,zander2025role,robbiatiDBQA2026,PhysRevLett.134.180602,son2025quantum,schwartzman2026imaginary,benavides2026augmenting,kolotouros2026quantum,ito2026quantum}.
%%%%%%%%%%%%%%%%%%%% Figure %%%%%%%%%%%%%%%%%%%% 
\begin{figure}[t]
    \includegraphics[width=\columnwidth]{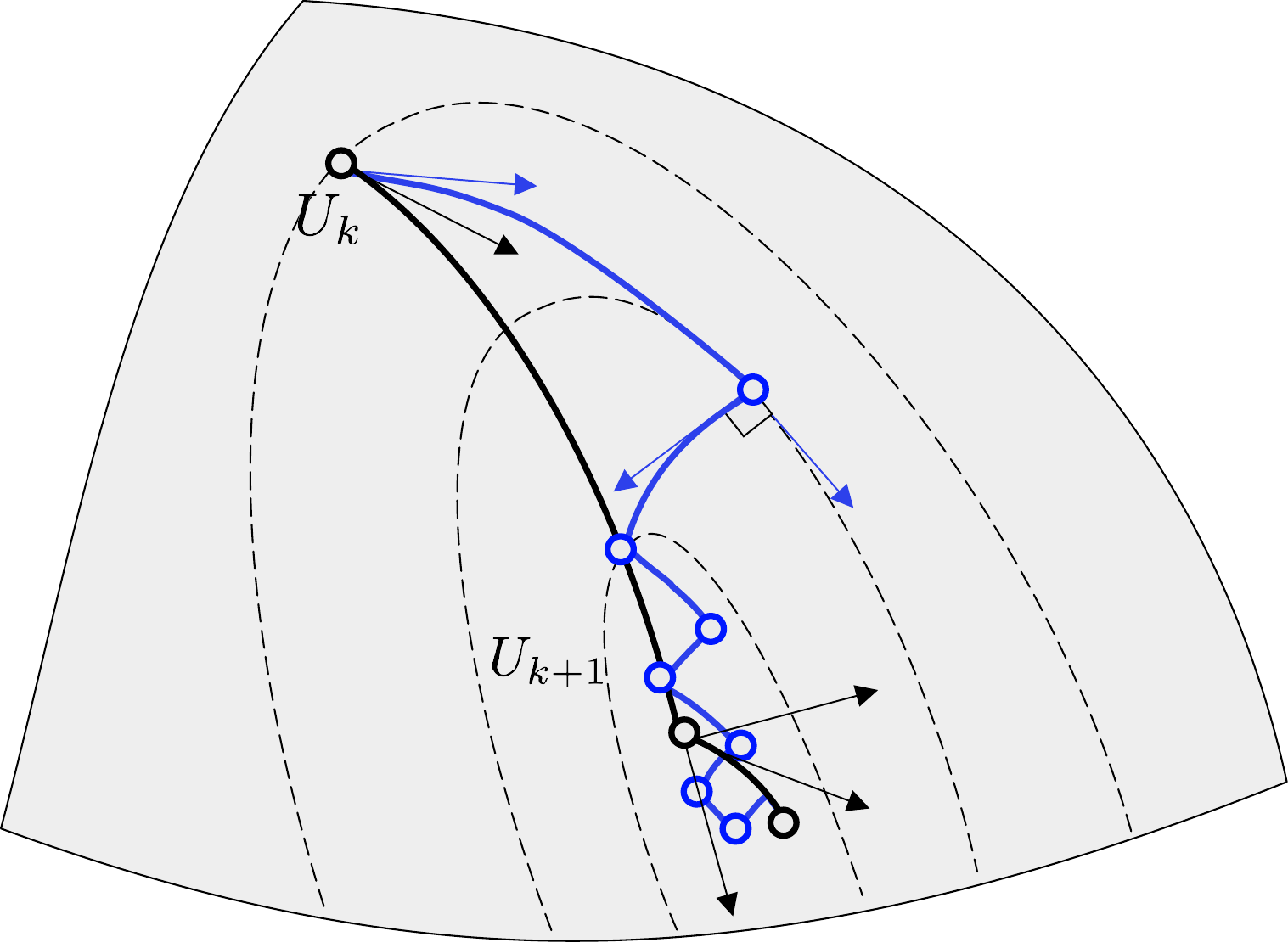}
    \caption{\textbf{Schematic comparison of conjugate-gradient descent (CGD; black) and gradient descent (GD; blue) for minimising a function \(E(U)\) on the unitary manifold.} Each curve segment represents a step along a geodesic. For GD with exact line search, the step size is chosen to minimise \(E(U)\) along the current geodesic; thus, the negative Riemannian gradient at the new iterate is orthogonal to the parallel transport of the previous search direction, producing the illustrated right-angle turns \cite{abrudanConjugateGradientAlgorithm2009, smith1993geometric}. At the final CGD iterate, the three arrows indicate, respectively, the parallel-transported previous CGD direction (tangent to the incoming path), the new negative Riemannian gradient (orthogonal to it under exact line search), and the resulting CGD search direction obtained by combining these two vectors. The first displayed CGD and GD steps do not coincide because we show the CGD trajectory after its initial gradient-descent step for conciseness.}
    \label{introduction.fig.GDvsCGD}
\end{figure}
%%%%%%%%%%%%%%%%%%%% Figure %%%%%%%%%%%%%%%%%%%%
Geometrically, we can view DB-QITE as a quantum implementation of Riemannian \emph{gradient descent} (GD) on the unitary manifold, minimising the energy cost function, \(E(U)\) defined later in Eq.~\eqref{eq:energy-cost-function}.  Algebraically, we have that DB-QITE is a quantum circuit approximating the minimisation process defined by
\begin{equation}
    \label{introduction.eq.recursion}
    U_{k+1} = e^{\mu_k  W_k} U_k,
\end{equation}
where \(U_k\) is the unitary preparing the \(k\)th approximation to the ground-state of a Hamiltonian \(H\), \(W_k\) is the \(k\)th gradient descent direction relative to \(E(U)\), and \(\mu_k\) a step size chosen to give an optimal energy reduction at step \(k\). DB-QITE can improve the fidelity to the ground-state exponentially, albeit at the expense of an exponential increase in circuit depth.

This perspective suggests that new GSP quantum algorithms may be obtained by using other \(W_k\) previously discussed in the literature on Riemannian optimization flows such as \emph{conjugate gradient descent} (CGD) and the Newton method \cite{laiAchievingDoublelogarithmicPrecision2026a}. This opens an opportunity to transfer decades of advances in manifold optimization \cite{boumalIntroductionOptimizationSmooth2023,AbsilMahonySepulchre+2009,huBriefIntroductionManifold2020} to quantum algorithm development. Indeed, several recent works have highlighted the importance of respecting the underlying geometry of quantum states not only in the GSP context \cite{ferreiramartins2026quantumoptimizationexactgeodesic,pervezRiemannianGradientDescentbased2025,malvettiRandomizedGradientDescents2025,magannRandomizedAdaptiveQuantum2023,laiQuantumCircuitDesign2026} but also in unstructured search \cite{suzukiGroversAlgorithmApproximation2026,laiGrovercompatibleManifoldOptimization2025,laiAchievingDoublelogarithmicPrecision2026a}, quantum tomography \cite{hsuQuantumStateTomography2024,liQuantumCombTomography2025}, Hamiltonian simulation \cite{leRiemannianQuantumCircuit2025,kotilRiemannianQuantumCircuit2024}, and circuit compilation \cite{yaoRiemannianOptimizationPhotonic2024,luchnikovRiemannianGeometryAutomatic2021a,luchnikovQGOptRiemannianOptimization2021}. Despite this promising connection, the role of memory-accelerated Riemannian methods remains largely unexplored within the quantum GSP context.

\begin{figure*}[t]
    \includegraphics[width=0.9\linewidth]{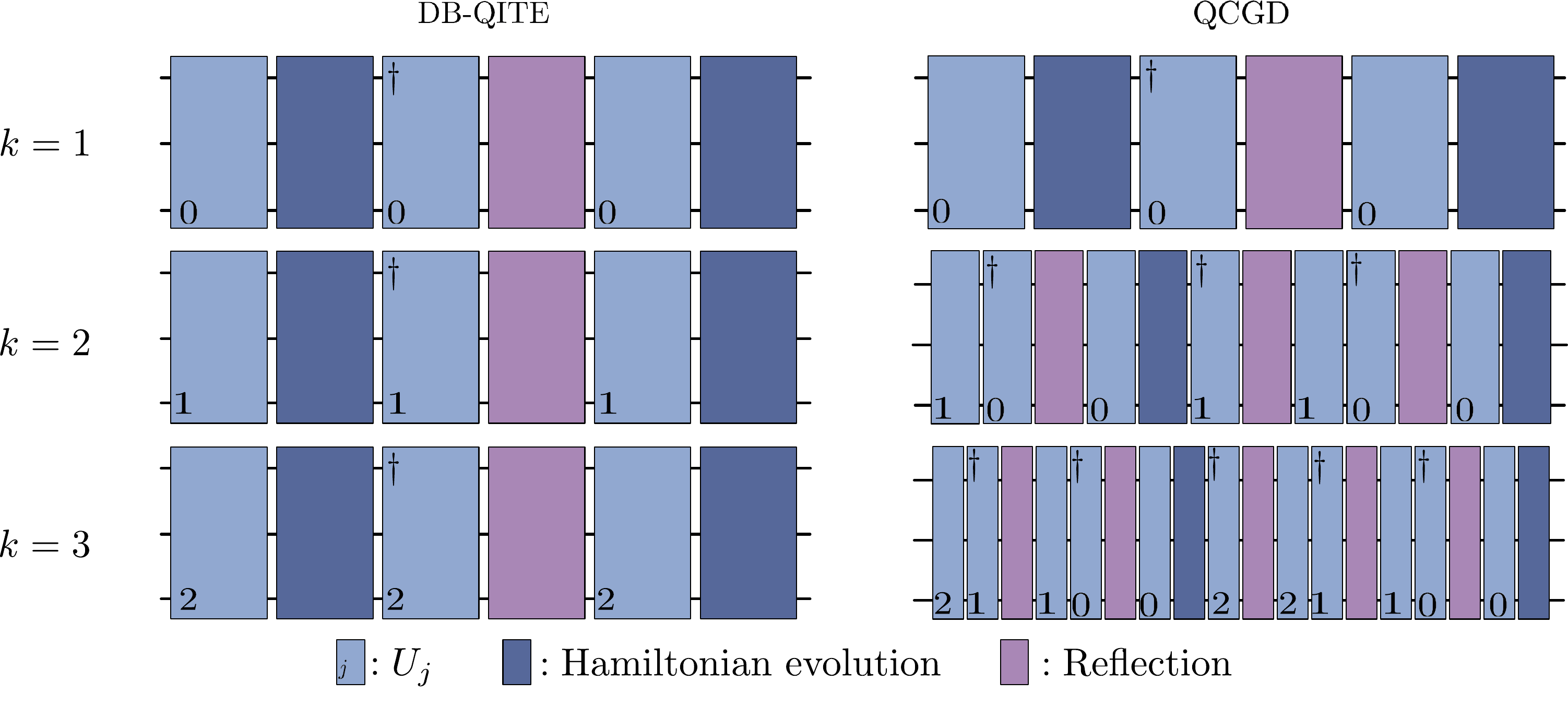}
    \caption{\textbf{Sketch of the DB-QITE and QCGD recursion rules.} Each row, labelled \(k\), represents the unitary update constructed at the \(k\)th iteration. The comparison shows how the QCGD circuit grows recursively as new operations are inserted. The light blue blocks denote the unitaries \(U_j\) that prepare approximations to the ground-state of a Hamiltonian \(H\), while the dark blue and purple blocks denote Hamiltonian evolution \(e^{it H}\) and generalised reflections \(e^{i t \ket{0}\bra{0}}\) about the \(\ket{0}\) state, respectively. The operations are applied from left to right, and \(\dagger\) indicates that the circuit should be inverted.}
 \label{introduction.fig.algorithm-sketch}
\end{figure*}

Classical CGD methods are among the most successful first-order techniques in classical optimization and, as illustrated in Fig.~\ref{introduction.fig.GDvsCGD}, they can achieve substantially better convergence while only slightly increasing computational cost \cite{nocedalNumericalOptimization1999,abrudanConjugateGradientAlgorithm2009}. In this work, we leverage the connection between DB-QITE and GD to enhance the performance of GSP with a memory effect inspired by conjugate gradient descent (CGD). As sketched in Fig.~\ref{introduction.fig.algorithm-sketch}, we develop a recursive quantum algorithm, labelled \emph{quantum conjugate gradient descent} (QCGD), that approximates the Riemannian CGD iteration that minimises \(E(U)\). Through numerical simulations,  we compare DB-QITE and QCGD updates and identify the trade-off between improved per-step energy reduction and increased circuit depth. 
Crucially, at each iterate, the QCGD candidate set contains the corresponding DB-QITE update. Thus, QCGD can lower a state's energy at least as much as the corresponding DB-QITE update. We expect the proposed approach to be particularly advantageous as a warm-start strategy in the early fault-tolerant era, or in synergy with more established \cite{nearoptimalground_Lin_2020,fragmented_QITE2024,dong2022ground,gilyen2019quantum,ge2019faster} and heuristic \cite{avqite,mcArdle2019,nishi2021,motta2020determining,yeter2020practical,MottaAdiabatic,KishorRevModPhys.94.015004,cerezo2021variational} GSP methods.

\section{Preliminaries}
\label{part1}

\subsection{Geometric Formulation for ground-state Preparation}
\label{geometric-formulation-for-ground-state-preparation}
Consider the problem of lowering the energy of an initial state, \(|0\rangle\), with respect to a Hamiltonian $H$. Let $\rho_0 = |0\rangle\langle 0|$ and $U\in \mathcal{U}(N)$ be a unitary operator, where \(\mathcal{U}(N)\) is the unitary group and $N=2^L$ for an $L$-qubit system. The state prepared by $U$ is 
\begin{align} 
    |\psi\rangle = U |0\rangle
\end{align}
and its energy is given by 
\begin{align}
\label{eq:energy-cost-function}
    E(U)=\langle 0|U^\dagger H U|0\rangle =\operatorname{Tr}\left(HU\rho_0U^\dagger\right) = \operatorname{Tr}\left( H\psi \right),
\end{align} 
where \(\psi = U \rho_0 U^\dagger\).

From a geometric perspective, we can formulate a GSP problem as the minimisation of $E: \mathcal{U}(N)\to \mathbb{R}$, where each step is taken along geodesics connecting the two points on the manifold \(U_k\) and \(U_{k+1}\). To identify a descent direction for such schemes, consider a geodesic on the unitary group emanating from an initial point \(U_0\). This is given by
\begin{equation}
    \label{part1.geodesic-update}
    U(t) = e^{tW}U_0,
\end{equation}
where \(W \in \mathfrak{u}(N) = \{M\in \mathbb{C}^{N \times N}: M^\dagger = -M\}\) is an element of the Lie algebra of $\mathcal{U}(N)$ and specifies the initial direction of the geodesic. Let us also define the tangent space at \(U\in \mathcal{U}(N)\) as 
\begin{align}
    T_U\mathcal{U}(N) = \{X \in \mathbb{C}^{N\times N}: UX^\dagger+ XU^\dagger  = 0\}.
\end{align}
Note that we can write any \(\widetilde{W}\in T_U\mathcal{U}(N)\) as \(\widetilde{W} = WU\), where \(W \in \mathfrak{u}(N)\), and that \(\mathfrak{u}(N)= T_I\mathcal{U}(N)\) is the tangent space at the identity \(I\in \mathcal{U}(N)\). We remark that unitary operators that differ only by a global phase are physically equivalent in quantum computing, and thus one may work with \(\mathrm{SU}(N)\) instead of \(\mathcal{U}(N)\).

Let \(\mathcal{M}\) be a Riemannian manifold with metric \(\langle\cdot,\cdot\rangle\). The Riemannian gradient of a function \(f:\mathcal{M}\rightarrow\mathbb{R}\) at \(x\in\mathcal{M}\) is the unique tangent vector \(\operatorname{grad}f(x)\in T_x\mathcal{M}\) satisfying
\begin{align}
    \left\langle\operatorname{grad}f(x),\eta\right\rangle_x= Df(x)[\eta],
\end{align}
for every tangent vector \(\eta\in T_x\mathcal{M}\). Here, \(\eta\) specifies a direction at \(x\), \(Df(x):T_x\mathcal{M}\rightarrow\mathbb{R}\) is the differential of \(f\) at \(x\), and \(Df(x)[\eta]\) is the directional derivative of \(f\) along \(\eta\). In the present setting, choosing \(\widetilde{W} = - \operatorname{grad}E(U)\) gives the GD minimisation scheme used in this work. Further details are provided in \cite{abrudanConjugateGradientAlgorithm2009,satoRiemannianConjugateGradient2022b}. 

The computation for \(\textrm{grad }E(U_0)\) follows from examining the change in energy along a general direction \(W,\)
\begin{equation}
    \label{part1.eq.cost-change}
    \Delta_W E(U(t)) = \frac{E\left( e^{tW}U_0\right) - E(U_0)}{t}.
\end{equation}
Following the details in Appendix \ref{riemannian gradient computation}, the Riemannian gradient is found to be \(\operatorname{grad} E\left(U_0\right)= [ H, \psi_0]U_0 \in T_{U_0}\mathcal{U}(N)\), where \([H, \psi_0] \in \mathfrak{u}(N)\) \cite{helmkeOptimizationDynamicalSystems1994,mooreNumericalGradientAlgorithms1994,schulte-herbruggenOptimalControlGenerating2011,wiersemaOptimizingQuantumCircuits2023b} . Choosing \(W = - [H,\psi_0]\) gives
\begin{equation}
    \left.\frac{d}{dt} E(U(t))\right|_{t=0} = - || [H, \psi_0]||^2_{\textrm{HS}} \leq 0.
\end{equation}
Thus, it yields the steepest local energy decrease in the Hilbert-Schmidt geometry. This motivates the GD iteration
\begin{equation}
\label{part1.eq.GD-iteration}
    U_{k+1}=e^{-\mu_k[H,\psi_k]}U_k,  \qquad |\psi_k \rangle=U_k|0\rangle,
\end{equation}
 where $\mu_k>0$ is selected by a line search to minimise $E(U_{k+1})$ and we define \(\psi_k =  U_k \rho_0 U_k^\dagger\). Results from \cite{helmkeOptimizationDynamicalSystems1994,mooreNumericalGradientAlgorithms1994,bloch1990steepest,smith1993geometric} show that as \(k\to\infty\), this process provably converges to the ground-state under appropriate assumptions on the initial state and step-size selection.

Although the update above is unitary by construction, the exponential \(e^{-\mu_k[H,\psi_k]}\) is still challenging to execute directly on a quantum computer. DB-QITE resolves this issue by using the group commutator \cite{dawson2006solovay,gluza2024doublebracket,chenEfficientProductFormulas2022} for two Hermitian operators \(\hat{A}, \hat{B}\)
\begin{equation}
    \label{eq:group-commutator}
     G_s(\hat A , \hat B ) = e^{i\sqrt{s}\hat A}e^{i\sqrt{s}\hat B}    
    e^{-i\sqrt{s}\hat A}
    e^{-i\sqrt{s}\hat B},
\end{equation}
to approximate \( e^{-s[\hat A, \hat B]} = G_s(\hat{A},\hat{B}) + O(s^{3/2})\) . As we show later, this also inspires our approach to QCGD.

\subsection{Double-Bracket Quantum Imaginary-Time Evolution}
\label{dbqite}
DB-QITE \cite{gluza_DB_QITE_2024} tackles the open problem of designing efficient quantum circuits guided by \emph{imaginary-time evolution} (ITE). Specifically, these are circuits which approximate the ground-state \(|\lambda_0\rangle\) of a Hamiltonian \(H\) by evolving an initial state \(|\Psi_0\rangle\) as 
\begin{equation}
    \label{introduction.eq.ITE}
    |\Psi(\tau) \rangle = \frac{e^{-\tau H} |\Psi_0\rangle}{||e^{-\tau H} |\Psi_0\rangle||},
\end{equation}
with \(|||\Omega\rangle|| = \sqrt{\langle \Omega | \Omega \rangle}\) for any vector \(|\Omega\rangle\). As \(\tau\to\infty\), \(|\Psi(\tau)\rangle\) is guaranteed to converge to \(|\lambda_0\rangle\) provided that \(|\Psi_0\rangle\) has non-zero overlap with the ground-state, that is \(\langle \lambda_0 | \Psi_0 \rangle \neq 0\)~\cite{GellmannLow}, and \(\ket{\lambda_0}\) is non-degenerate.

Unlike quantum circuits, ITE is a non-unitary process. Thus, the fundamental algorithmic-design task lies in finding unitaries \(Q_\tau\) such that \(|\Psi(\tau) \rangle \approx Q_\tau |\Psi_0 \rangle \), giving rise to the notion of \emph{quantum imaginary-time evolution} (QITE). DB-QITE exploits the relation between ITE and \emph{double bracket flows}, together with the \emph{double bracket quantum algorithm} framework~\cite{gluza2024doublebracket,robbiatiDBQA2026}, to propose a non-variational, systematic approach with potential applications to near-term quantum hardware.

From a geometric perspective, we can understand DB-QITE as a quantum-compatible implementation of the GD iteration defined in Eq.~\eqref{part1.eq.GD-iteration}. More broadly, we call a minimisation scheme quantum compatible when its unitary update is well suited for implementation on a quantum computer. For DB-QITE, this requires approximating the commutator exponential appearing in the GD update using Hamiltonian evolutions \(e^{i\alpha H}\) and generalised phase rotations \(e^{i\beta\psi_j}\), which can be implemented using the recursively available unitaries \(U_j\) and \(U_j^\dagger\).

\begin{figure}
    \includegraphics[width=0.5\columnwidth]{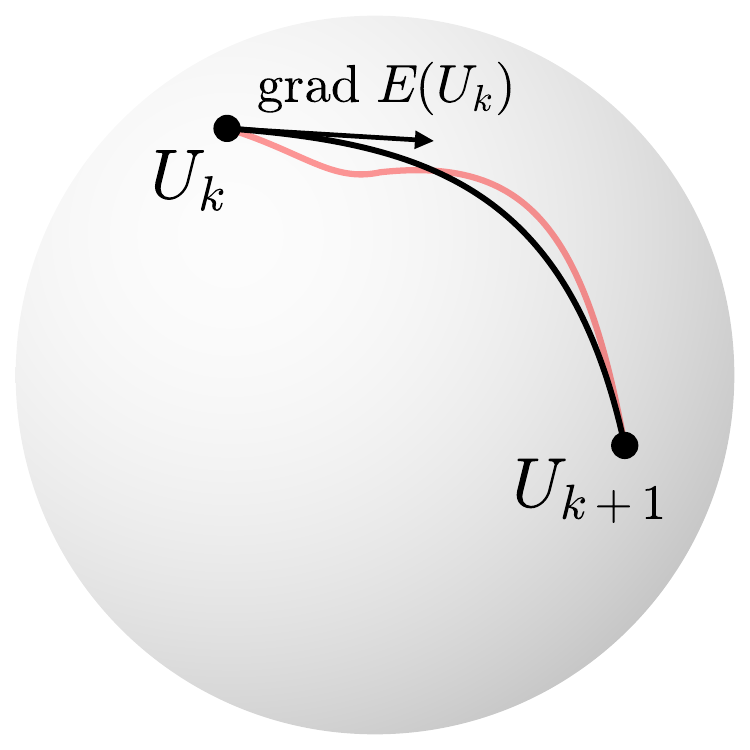}
    \caption{\textbf{Schematic comparison between the exact gradient-descent geodesic (black) and its group-commutator approximation (red).}  The exact path is generated by \(e^{-\mu[H,\psi_k]}U_k\), while the product-formula path is constructed from Hamiltonian evolutions \(e^{i\alpha H}\) and phase rotations \(e^{i\beta\psi_k}\).}
    \label{part2.fig.geodesic-update}
\end{figure}

Following DB-QITE's approach, we employ the group commutator product formula, proposed in Ref.~\cite{gluza2024doublebracket}, to approximate the commutator exponential in Eq.~\eqref{part1.eq.GD-iteration} as
\begin{align}
    \label{part2.eq.product-formula-3/2}
    e^{-\mu [H, \psi_k]} = G_\mu(H,\psi_k)+ O\left(\mu^{3/2}\right).
\end{align}
The final factor \(e^{-i\sqrt{\mu_k}\psi_k}\) from the group commutator only contributes a global phase when acting on \(|\psi_k\rangle\). Consequently, this factor may be omitted without affecting the resulting quantum state, yielding the recursive state-preparation formula
\begin{equation}
 \label{part2.eq.GCI-proposition}
    U_{k+1} = e^{i \sqrt{\mu_k} H} U_k e^{i \sqrt{\mu_k} \rho_0}U_k^{\dagger} e^{-i\sqrt{\mu_k} H} U_k.
\end{equation}
At each step, we choose \(\mu_k\) via a one-dimensional line search over a fixed interval, as in \cite{gluza_DB_QITE_2024}, yielding Algorithm 1. This recovers the DB-QITE update rule as outlined in Fig.~\ref{introduction.fig.algorithm-sketch} and sketched pictorially in Fig.~\ref{part2.fig.geodesic-update}. This scheme provides strong energy reductions and fidelity guarantees, as shown in \cite{gluza_DB_QITE_2024}. In Appendix \ref{dbqite hopf}, we also provide details of higher-order product formulae which reduce the error in Eq.~\eqref{part2.eq.product-formula-3/2} to \(O\left(\mu^2\right)\).

\begin{algorithm}[H]
\caption{\(\textrm{DB-QITE}\)}
\begin{algorithmic}[1]
\State Initialise \(U_0\)
\State \(\ket{\psi_0} \gets U_0\ket{0}\)
\State \(k \gets 0\)
\State Initialise \(\textrm{max}_k\)
\While{\(k  < \textrm{max}_k\)}
    \State  \(\mu_k \gets \arg\min_{\mu} E\!\left(\textrm{DB-QITE}(\mu)U_k\right)\)
    \State \(U_{k+1} \gets \textrm{DB-QITE}(\mu_k)U_k\)
    \State \(\ket{\psi_{k+1}} \gets U_{k+1}\ket{0}\)
    \State \(k \gets k+1\)
\EndWhile
\end{algorithmic}
\end{algorithm}
As demonstrated in more detail in Appendix \ref{dbqite query complexity}, the query complexity of DB-QITE scales as \(O\left(3^k\right)\),  making the number of iterations a key factor in determining the computational cost. Thus, this work aims to construct a memory-augmented search direction that produces an energy reduction large enough to offset the exponential circuit growth, while remaining expressible in the same quantum-compatible primitives.

\subsection{Overview of Riemannian Conjugate Gradient Descent}
\label{part2}
Originally, CGD was proposed to solve linear equations of the form \(Ax = b\) where \(A \in \mathbb{R}^{n \times n}\) is symmetric and positive definite, and \(b \in \mathbb{R}^n\) \cite{hestenesMethodsConjugateGradients}. This method can be reformulated to minimise strictly convex quadratic functions \(f(x) = x^TAx/2 - b^T x\) for \(x \in \mathbb{R}^n\) by finding the minimum point \(\hat{x}\) of \(f\), where \(\hat{x} = A^{-1}b\) since \(\nabla f(x) = Ax-b\).  Moreover, this extends to more general optimization problems involving a sufficiently smooth cost function \cite{fletcherFunctionMinimizationConjugate1964}. Given a cost \(f: \mathbb{R}^n \to \mathbb{R}\), CGD generates a sequence of points \(\{x_k\}\) in \(\mathbb{R}^n\), starting from an initial  point \(x_0\), given by 
\begin{equation}
\label{part3.eq.flat-CGD-line-search}
    x_{k+1} = x_k + t_k \eta_k, \quad k \geq 0.
\end{equation}
Here, \(\eta_k \in \mathbb{R}^n\) is the search direction and is computed as \(\eta_0 = - \nabla f(x_0)\) and 
\begin{equation}
\label{part3.eq.flat-CGD}
    \eta_{k+1} = - \nabla f(x_{k+1}) + \gamma_{k+1} \eta_k,
\end{equation}
for \(k\geq0\), where \(\gamma_{k+1} \in \mathbb{R}\) and \(t_k>0\) are chosen to satisfy certain criteria as detailed in \cite{satoRiemannianConjugateGradient2022b}. By introducing a memory of previous search directions, CGD avoids the zig-zag behaviour characteristic of GD, as sketched in Fig.~\ref{introduction.fig.GDvsCGD}, and, under the right conditions, can converge faster to a minimum.

In general,  this method can be extended to any Riemannian manifold \(\mathcal{M}\) \cite{satoRiemannianConjugateGradient2022b} by appropriately generalising the line search strategy in Eq.~\eqref{part3.eq.flat-CGD-line-search} and Eq.~\eqref{part3.eq.flat-CGD}. First, we must choose a retraction map \(R_x: T_x\mathcal{M}\to \mathcal{M}\) to generate the generalised iteration sequence 
\begin{equation}
    x_{k+1} = R_{x_k}(t_k\eta_k).
\end{equation}
Then, we upgrade the term \(\nabla f(x)\) to its Riemannian equivalent, \(\textrm{grad }f(x)\), and adjust Eq.~\eqref{part3.eq.flat-CGD} so that both terms can be combined appropriately. Under standard assumptions on the line search and vector transport, several Riemannian conjugate-gradient variants have been shown to converge to a stationary point under relevant assumptions. However, they do not guarantee a better worst-case convergence rate than Riemannian gradient descent \cite{satoRiemannianConjugateGradient2022b,zhouNOVELRIEMANNIANCONJUGATE,boumalGlobalRatesConvergence2019}.

Regarding our specific optimization problem outlined in Section \ref{part1}, we proceed in a similar fashion to \cite{abrudanConjugateGradientAlgorithm2009} and employ the geodesic search
\begin{equation}
\label{eq: general unitary update}
U_{k+1} = e^{\mu_kW_k}U_k, \quad k \geq 0 ,
\end{equation}
where we use the exponential function as our retraction map, $R_{U}(\mu \widetilde{W})= \exp(\mu \widetilde{W}U^\dagger)U$. Here, \(U_k \in \mathcal{U}(N)\) and we relabel the search direction \(\eta_k\) as \(\widetilde{W}_k \in T_{U_k}\mathcal{U}(N)\), such that \(\widetilde{W_k} = W_k U_k\), where \(W_k \in \mathfrak{u}(N)\). The initial search direction now becomes \(W_0 = - [H,\psi_0]\) and the first term on the right-hand side of Eq.~\eqref{part3.eq.flat-CGD} becomes \(- \textrm{grad } E(U_{k+1}) \in T_{U_{k+1}}\mathcal{U}(N)\). Since the two terms on the right-hand side belong to distinct tangent spaces, we cannot add them directly. 
\begin{figure*}[t]
    \centering
    \includegraphics[width=\linewidth]{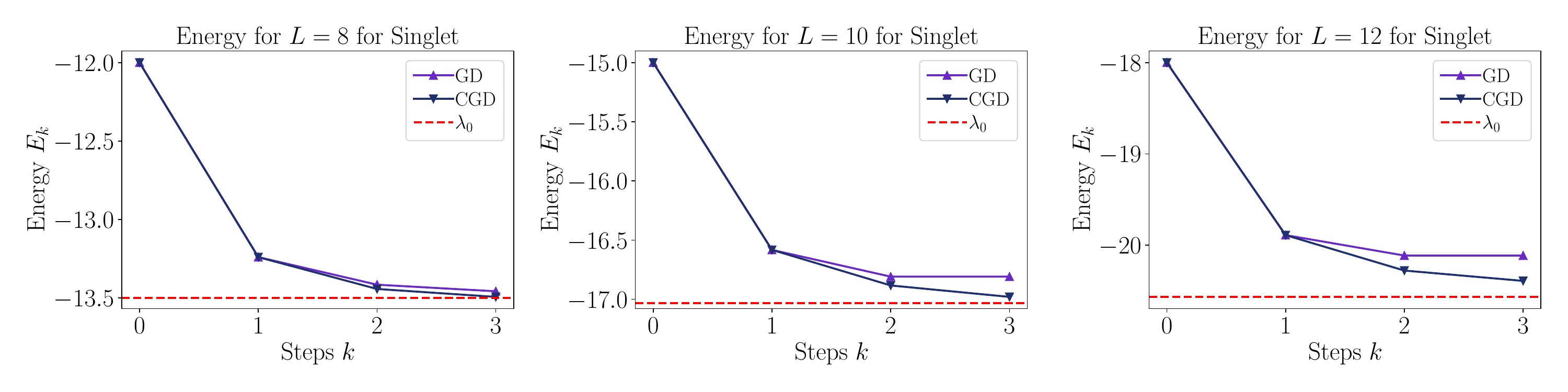}
    \caption{\textbf{Comparison of Riemannian gradient descent (GD) and conjugate-gradient descent (CGD) for the Heisenberg Hamiltonian introduced in Eq.~\eqref{eq: Heisenberg Hamiltonian} with the singlet initialisation from Eq.~\eqref{eq: singlet}.} From left to right, the panels show the energy \(E_k=\langle\psi_k|H|\psi_k\rangle\) after optimization step \(k\) for \(L=8\), \(10\), and \(12\) qubits. The red dashed line marks the exact ground-state energy, \(\lambda_0\), for each system size. The two methods coincide at the first update because CGD has no previous search direction. From the second update onward, CGD uses information retained from the preceding step and consistently reaches a lower energy than GD. The improvement is modest for \(L=8\) but becomes increasingly visible for \(L=10\) and \(12\), where GD begins to plateau while CGD continues to reduce the energy. The figure illustrates the central practical benefit of the conjugate-gradient direction which we hope to retain when implementing its quantum-compatible equivalent.}
    \label{fig: cgd vs gd}
\end{figure*}
Hence, to ensure compatibility, we appropriately transport \cite{satoRiemannianConjugateGradient2022b} \(\widetilde{W}_k\) from \(T_{U_k}\mathcal{U}(N)\) to \(T_{U_{k+1}}\mathcal{U}(N)\),  and denote this as \(\Gamma_k^{k+1}\widetilde{W_k}\). One such valid transport is to parallel transport \(\widetilde{W}_k= W_k U_k\) along the geodesic emanating from \(U_k\) in the direction of \(\widetilde{W}_k\), giving
\begin{align}
    \Gamma_k^{k+1} \widetilde{W}_k &= \widetilde{W}_k U_k^\dagger e^{tW_k}U_k \\ 
                                &= W_ke^{tW_k}U_k, \label{eq: parallael transport}
\end{align}
as described in \cite{abrudanConjugateGradientAlgorithm2009}. For \(t=\mu_k\), Eq.~\eqref{eq: general unitary update} implies that Eq.~\eqref{eq: parallael transport} reduces to  \(\Gamma_k^{k+1} \widetilde{W}_k = W_kU_{k+1}\), meaning that right multiplication by \(U_k^\dagger U_{k+1}\) is equivalent to the parallel transport of \(\widetilde{W}_k\) from \(T_{U_k}\mathcal{U}(N)\) to \(T_{U_{k+1}}\mathcal{U}(N)\) along the geodesic in the \(\widetilde{W}_k\) direction. This results in the following update for the search direction
\begin{equation}
    \widetilde{W}_{k+1} = - \textrm{grad } E(U_{k+1}) + \gamma_{k+1} W_kU_{k+1},
\end{equation}
which, after right multiplication by \(U_{k+1}^\dagger\), simplifies to 
\begin{equation}
\label{part3.eq.search-recursion}
    W_{k+1} = - [H,\psi_{k+1}] + \gamma_{k+1} W_k.
\end{equation}
This allows us to define our Riemannian CGD scheme as 
\begin{equation}
    \label{part3.eq.CGD-update}
    U_{k+1} = \begin{cases} e^{\mu_k(-[H, \psi_k] + \gamma_k W_{k-1})}U_k & k >0, \\
    e^{-\mu_0[H,\psi_0]}U_0 & k=0,
    \end{cases}
\end{equation}
where we choose to select \(\mu_k\) and \(\gamma_k\) over a fixed grid such that \(E(U_{k+1})\) is minimal. As explored in \cite{satoRiemannianConjugateGradient2022b}, different choices of transport and parameter selection methods for \(\mu_k,\gamma_k\) lead to different valid Riemannian CGD schemes. The motivation for our proposed quantum algorithm stems from the observation that CGD can achieve a larger per-step energy decrease than GD both in Euclidean space and on Riemannian manifolds such as \(\mathcal{U}(N)\) \cite{abrudanConjugateGradientAlgorithm2009}.
We illustrate this for the classical optimization in Fig.~\ref{fig: cgd vs gd}, suggesting that the quantum case can enjoy similar benefits.
 
\section{Main Results}

\subsection{Proposed Method: Quantum Conjugate Gradient Descent}
\label{part3}
The QCGD recursion derived below generates a sequence of unitaries \(U_k^{\mathrm{QCGD}}\) and associated states
\begin{align}
    \lvert\psi_k^{\mathrm{QCGD}}\rangle
    := U_k^{\mathrm{QCGD}}\lvert 0\rangle, \
    \psi_k^{\mathrm{QCGD}} := \lvert\psi_k^{\mathrm{QCGD}}\rangle
    \langle\psi_k^{\mathrm{QCGD}}\rvert.
    \label{eq:qcgd-states-label}
\end{align}
To simplify the notation, throughout this subsection we write \(U_k=U_k^{\mathrm{QCGD}}\) and \(\psi_k=\psi_k^{\mathrm{QCGD}}\).

We define QCGD by first writing the search direction in the form \(W_k = -[H, \phi_k]\). Expanding the recursion relation in Eq.~\eqref{part3.eq.search-recursion}, we obtain
\begin{align}
        W_k = -[H,\psi_k] &- \gamma_k[H, \psi_{k-1}] - \gamma_k \gamma_{k-1}[H, \psi_{k-2}] \nonumber \\ &-\dots - \gamma_k \cdots \gamma_1[H,\psi_0],
    \end{align}
where \(\psi_j = U_j |0\rangle \langle 0 | U_j^\dagger\) and \(\gamma_j\) has been chosen, together with \(\mu_j\), to minimise \(E(U_{j+1})\) for each \(j=0,\dots,k-1\). Thus, for \(k\geq 1\), \(\phi_k = \sum_{j=0}^k c_j \psi_j\) where
\begin{equation}
    c_j = \begin{cases} \prod_{i=j+1}^{k} \gamma_i & 0 \leq j \leq k-1 ,\\ 1 & j=k,\end{cases}
\end{equation}
and \(\phi_0 = \psi_0\).  Employing the group commutator as in Section \ref{dbqite} for \(H\) and \(\phi_k\) gives
\begin{equation}
    e^{-\mu_k[H,\phi_k]} = e^{i \sqrt{\mu_k} H} e^{i \sqrt{\mu_k} \phi_k} e^{-i \sqrt{\mu_k}H} e^{-i \sqrt{\mu_k} \phi_k} + O\left(\mu_k^{3/2}\right).
\end{equation}
Additionally, to ensure quantum compatibility, we approximate the exponentials in \(\phi_k\) via the Trotter-Suzuki formula \cite{childs2021theory}. In order to retain the group commutator structure \(G = QVQ^\dagger V^\dagger\) for respective unitaries \(Q\) and \(V\), we approximate
\begin{align}
    e^{i\sqrt{\mu_k}\phi_k} &\approx e^{i \sqrt{\mu_k}c_k \psi_k} \cdots e^{i \sqrt{\mu_k}c_0\psi_0}, \label{eq:reflection-sum-approx+} \\
    e^{-i\sqrt{\mu_k}\phi_k} & \approx e^{-i \sqrt{\mu_k}c_0 \psi_0} \cdots e^{-i \sqrt{\mu_k}c_k\psi_k}, \label{eq:reflection-sum-approx-}
\end{align}
leading to

\begin{align}
    \label{part3.eq.TGCI-3/2}
      U_{k+1} = &e^{i \sqrt{\mu_k}H} e^{i \sqrt{\mu_k} c_k \psi_k} \cdots e^{i\sqrt{\mu_k}c_0\psi_0} \nonumber \\ &e^{-i \sqrt{\mu_k}H} e^{-i\sqrt{\mu_k}c_0\psi_0}  \cdots e^{-i \sqrt{\mu_k} c_k \psi_k} U_k.
\end{align}
For numerical simplicity, we relabel  \(\gamma_k\)  as \(\gamma_k / \sqrt{\mu_k}\), resulting in the new phase coefficients 
\begin{equation}
\label{eq:modified-coefficients}
\theta_j = \sqrt{\mu_k}c_j = \begin{cases} \prod_{i=j+1}^{k} \gamma_i & 0 \leq j \leq k-1, \\ 1 & j=k,\end{cases}
\end{equation} 
where we adopt the convention \(\theta_j = 0\) at \(\mu_k =0\) for \(0\leq j\leq k\). Finally, the right-most operation acts on \(U_k\) and only contributes a global phase factor. Omitting this phase results in
\begin{align}
\label{eq: qcgd update}
      U_{k+1} = &e^{i \sqrt{\mu_k}H} e^{i  \sqrt{\mu_k} \psi_k} e^{i\theta_{k-1}\psi_{k-1}}  \cdots e^{i\theta_0\psi_0} \nonumber \\ &e^{-i \sqrt{\mu_k}H} e^{-i\theta_0\psi_0}  \cdots e^{-i \theta_{k-1} \psi_{k-1}} U_k,\quad k \geq 1,
\end{align}
where \(U_1\) is the same as DB-QITE at the first step. Alternatively, one could first apply the Trotter–Suzuki approximation to $\sum_j c_j[H,\psi_j]$ and subsequently construct the group commutator. This ordering requires two Hamiltonian-evolution calls for each projector \(\psi_j\), resulting in substantially more Hamiltonian-evolution calls. Defining QCGD as in Eq.~\eqref{eq: qcgd update} is consistent with the expected behaviour of Riemannian CGD at \(\mu_k=0\), as it also returns the identity for any \(\gamma_k\). This leads to Algorithm~\ref{qcgd algorithm}.

\begin{algorithm}[H]
\caption{\(\textrm{QCGD}\)}
\label{qcgd algorithm}
\begin{algorithmic}[1]
\State Initialise \(U_0\)
\State \(\ket{\psi_0} \gets U_0\ket{0}\)
\State \(k \gets 0\)
\State Initialise \(\textrm{max}_k\)
\While{\(k  < \textrm{max}_k\)}
    \State  \(\mu_k, \gamma_k \gets \arg\min_{\mu,\gamma} E\!\left(\textrm{QCGD}(\mu,\gamma)U_k\right)\)
    \State \(U_{k+1} \gets \textrm{QCGD}(\mu_k,\gamma_k)U_k\)
    \State \(\ket{\psi_{k+1}} \gets U_{k+1}\ket{0}\)
    \State \(k \gets k+1\)
\EndWhile
\end{algorithmic}
\end{algorithm}

By including \(\gamma_k=0\) in the QCGD line search, QCGD can always recover the corresponding DB-QITE update at that step and thereby inherit its guarantees of decreasing the energy. Therefore, we can choose the QCGD update so that its energy decrease is at least as large as that of the corresponding DB-QITE step.

As with DB-QITE, we provide details of a higher-order QCGD method in Appendix \ref{qcgd hopf} and of the associated query complexity in Appendix \ref{qcgd query complexity}.

\subsection{Numerical Setting}
\label{part4}
We numerically investigate whether QCGD's conjugate-gradient property yields a practically meaningful improvement over DB-QITE. In particular, we examine the following three considerations. First, in Section \ref{conjugate contribution}, we explore whether the optimized QCGD update selects a nonzero conjugate contribution and thereby improves upon the DB-QITE direction. Second, in Section \ref{energy and fidelity}, we ask if QCGD improves energy convergence and ground-state fidelity across system sizes and initial states. Third, in Section \ref{circuit resources}, we investigate if QCGD's faster convergence can offset its higher circuit cost compared with DB-QITE at matched fidelity. In summary, Fig.~\ref{fig: representative data} and Fig.~\ref{fig: energies and fidelities across L} show that QCGD consistently improves energy reduction and ground-state fidelity across the tested trajectories. Moreover, although QCGD is more expensive than DB-QITE for a fixed iteration count, Fig.~\ref{fig: circuit data} shows that its improved search direction matches or exceeds the fidelity obtained by DB-QITE with shorter circuits.

We consider the antiferromagnetic Heisenberg Hamiltonian
\begin{equation}
\label{eq: Heisenberg Hamiltonian}
    H = \sum_{i=1}^{L-1} \left(X_{i}  X_{i+1} + Y_iY_{i+1} + Z_i Z_{i+1}\right),
\end{equation}
for \(L \in \{10,12,14,16,18,20\}\). We denote the corresponding ground-state and first-excited-state energies by \(\lambda_0\) and \(\lambda_1\) respectively.

As in Ref.~\cite{gluza_DB_QITE_2024}, we use the Qrisp programming framework \cite{qrisp} and its sparse state-vector simulator to simulate the circuits implementing the unitary updates generated by DB-QITE and QCGD. Hamiltonian evolutions \(e^{i\tau H}\) are implemented using a second-order Trotter-Suzuki formula with two Trotter steps. Both algorithms use the same Hamiltonian simulation and compilation settings, so performance differences arise from the update rules and selected parameters rather than different simulation backends. The code used to perform the simulations and generate the results is publicly available in \cite{ArciaLopez2026QCGD}.

At iteration \(k\), DB-QITE selects the step size \(\mu_k\) by minimising the post-update energy over 20 equally spaced values in \([0,0.5]\), see Ref.~\cite{xiaoyue2024strategies} for more discussion. QCGD instead selects 
\begin{equation}
    (\mu_k,\gamma_k) = \mathrm{arg min}_{(\mu,\gamma) \in \mathcal{G}} E(QCGD(\mu,\gamma)U_k),
\end{equation}
where \(\mathcal{G}\) is the Cartesian product of 20 equally spaced values in \([0,0.5]\) for \(\mu_k\) and \(\gamma_k\), respectively. Thus, DB-QITE evaluates 20 candidate updates per iteration, while QCGD evaluates 400. The resource counts reported below describe the selected state preparation circuit and do not include this search overhead. We return to this distinction in Section \ref{limitations}.

As in \cite{gluza_DB_QITE_2024}, we use two physically inspired initialisations across our simulations. The first is the product of nearest-neighbour singlets
\begin{equation}
\label{eq: singlet}
    \ket{\text{Singlet}} = 2^{-L/4}(\ket{10}-\ket{01})^{\otimes L/2},
\end{equation}
which has an analytic preparation circuit and remains well-defined as the system size increases. The second is based on a variational circuit warm start applied on top of the singlet initialisation, called the Hamiltonian Variational Ansatz (HVA) \cite{bosse_Heisenberg_2022, kattemolle_vanwezel_2022,wecker2015progress}
\begin{equation}
\ket{\text{HVA}_p(\mathbf{t})} = \prod_{j=1}^{p}(e^{-it_{j,0} H_0}e^{-it_{j,1} H_1})\ket{\text{Singlet}}.
\end{equation}
This approach provides a more accurate approximation to the ground-state, but its scalability may be limited. Here, \(H_0, H_1\) are obtained from $H$ by restricting to summation over odd (even) indices $i$ such that the terms witihin each of $H_0$ and $H_1$ commute. In our numerics, we use just one layer, $p=1$, because it is enough to illustrate a significant improvement in the initialisation. Explicitly, the second initialisation is given by
\begin{align}
\ket{\text{HVA}} &= e^{-it_{0} H_0}e^{-it_{1} H_1}\ket{\text{Singlet}}\\ 
&= U_\text{HVA}\ket{\text{Singlet}}.
\end{align}

\subsection{Numerical Results}
\label{numerical results}
\begin{figure*}[t]
    \centering
        \includegraphics[width=\linewidth]{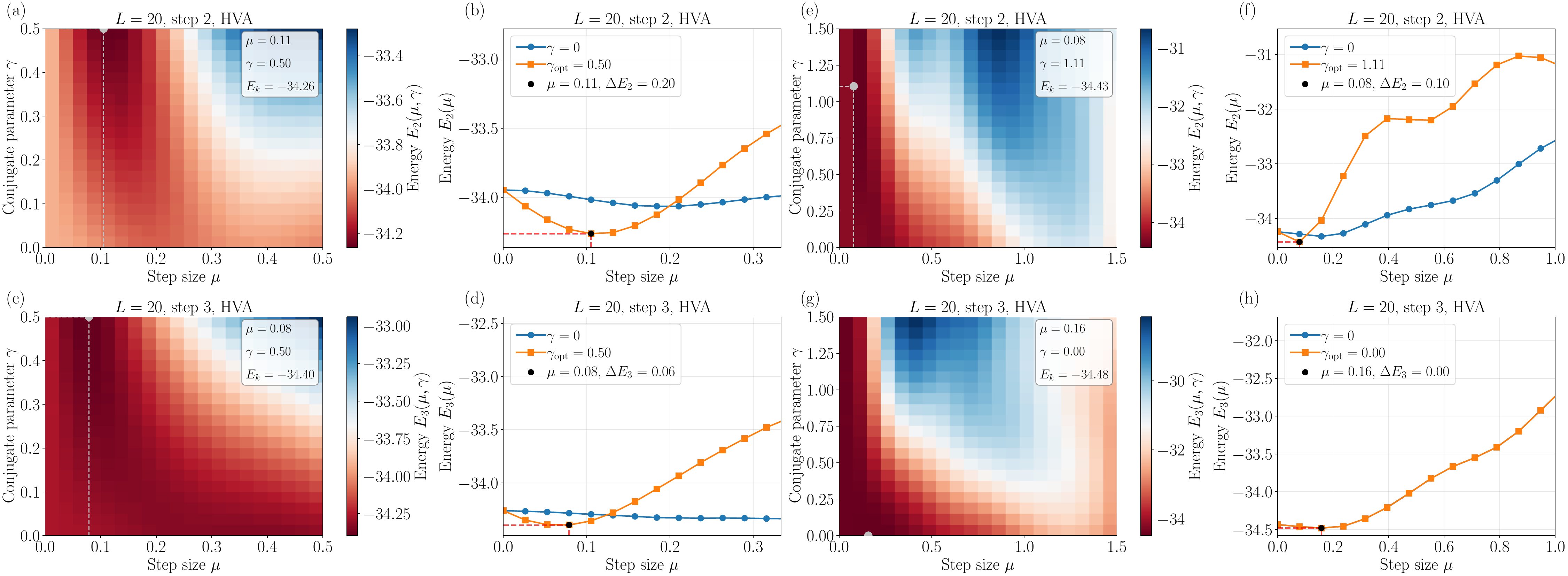}
    \caption{\textbf{QCGD parameter landscapes for the \(L=20\) Heisenberg model with HVA initialisation.} Panels (a)--(d), occupying the left half of the figure, show results for \((\mu,\gamma)\in[0,0.5]^2\); panels (e)--(h), in the right half, show results for the extended domain \([0,1.5]^2\). The upper and lower rows correspond to the second and third updates, respectively. Within each domain, the left panels show \(E_k(\mu,\gamma)\), with silver markers denoting the minima on the sampled grids, while the right panels compare the DB-QITE limit \(\gamma=0\) with the slices at the selected \(\gamma_{\textrm{opt}}\). Black markers indicate the optimal \(\mu\), and \(\Delta E_k\) denotes the energy reduction relative to DB-QITE. In the restricted domain, both minima occur at the boundary \(\gamma=0.5\). Extending the domain yields \(\gamma_{\textrm{opt}}=1.11\) at the second update and \(\gamma_{\textrm{opt}}=0\) at the third, illustrating the sensitivity of a finite-grid search to its parameter range.}
    \label{fig: heatmaps}
\end{figure*}
\subsubsection{Contribution of the conjugate direction}
\label{conjugate contribution}
We first check whether selecting a non-zero \(\gamma_k\)  always improves upon the DB-QITE direction. Note that the QCGD search domain contains \(\gamma_k=0\). For this value, the memory terms vanish, and the update reduces to the DB-QITE direction evaluated at the current iterate. Therefore, an exact minimisation over the specified domain cannot select an update with a higher energy than this fallback. Hence, a strictly lower energy than the best \(\gamma_k =0 \) candidate demonstrates a benefit from the memory contribution.

Fig.~\ref{fig: heatmaps} shows the QCGD energy landscapes for the \(L=20\) Heisenberg model with HVA initialisation. Over the restricted domain \((\mu,\gamma)\in[0,0.5]^2\), the energy depends non-trivially on both parameters, and the sampled minima for the second and third updates occur at the boundary \(\gamma=0.5\). This indicates that the restricted search is limited by its range and that the selected parameters do not approximate the minima of the broader landscape. Extending the domain to \([0,1.5]^2\) selects a nonzero conjugate contribution, \(\gamma_{\mathrm{opt}}=1.11\), at the second update, but \(\gamma_{\mathrm{opt}}=0\) at the third. With this extended domain, QCGD achieves a lower energy at each iteration, demonstrating that the energy improvement accumulated in an earlier update can be retained even when a subsequent update selects \(\gamma=0\). Thus, the numerical evidence supports an adaptive advantage, rather than the stronger claim that the conjugate contribution must be nonzero at every iteration.

\subsubsection{Energy and ground-state fidelity across system sizes}
\label{energy and fidelity}

For \(\mathcal A\in\{\mathrm{QCGD},\mathrm{DB\mbox{-}QITE}\}\), let \(\lvert\psi_k^{\mathcal A}\rangle\) denote the state obtained after \(k\) updates of algorithm \(\mathcal A\), with the QCGD states defined in Eq.~\eqref{eq:qcgd-states-label}. We denote the corresponding energy by
\begin{align}
    E_k^{\mathcal A}
    :=\langle\psi_k^{\mathcal A}\rvert
       H
       \lvert\psi_k^{\mathcal A}\rangle ,
    \label{eq:algorithmenergy}
\end{align}
and when the algorithm can clearly be inferred from context, we suppress the superscript \(\mathcal A\). To compare the energy decrease across system sizes, we report the normalised excess energy
\begin{align}
    r_k^{\mathcal A}
    =\frac{E_k^{\mathcal A}-\lambda_0}
           {E_0-\lambda_0},
\end{align}
along with the percentage of the initial excess energy removed by iteration \(k\),
\begin{equation}
    R_k^{\mathcal{A}} = 100(1-r_k^{\mathcal{A}}).
\end{equation}
Clearly, \(r_0^{\mathcal{A}} =1\), while \(r_k^{\mathcal{A}}=0\) corresponds to the exact ground-state energy. We quantify the QCGD advantage by the percentage difference
\begin{equation}
    \Delta R_k = R_k^{\text{QCGD}} - R_k^{\text{DB-QITE}}.
\end{equation}
Because they start with the same initial update, the first QCGD and DB-QITE steps coincide, and their trajectories diverge in subsequent iterations. Fig.~\ref{fig: representative data} shows representative absolute energy trajectories, the normalised excess energy, and the resulting fidelity as a function of the \(\mathrm{CZ}\) count. For \(L=20\) with HVA initialisation, QCGD produces a larger energy reduction at the second update and retains this accumulated advantage. For the singlet initialisation, the trajectory for \(L=14\) provides the largest tested system size for which QCGD reaches \(E_k < \lambda_1\) within three updates.

From Table~\ref{tab:comparison}, we observe that across all six system sizes, the singlet initialisation gives a particularly regular improvement. Explicitly \(\Delta R_2\) remains between \(23.0\%\) and \(24.1\%\), while \(\Delta R_3\) ranges from \(22.6\%\) to \(30.5\%\). QCGD also improves the optimization trajectories starting from the HVA initial state across all tested system sizes, although the extent of the improvement is more instance-dependent. In particular, the advantage for the HVA trajectory for \(L=14\) is smaller than for the other system sizes. This suggests that QCGD is sensitive not only to system size but also to the initial optimization landscape.

\begin{figure*}
    \centering
    \includegraphics[scale=0.35]{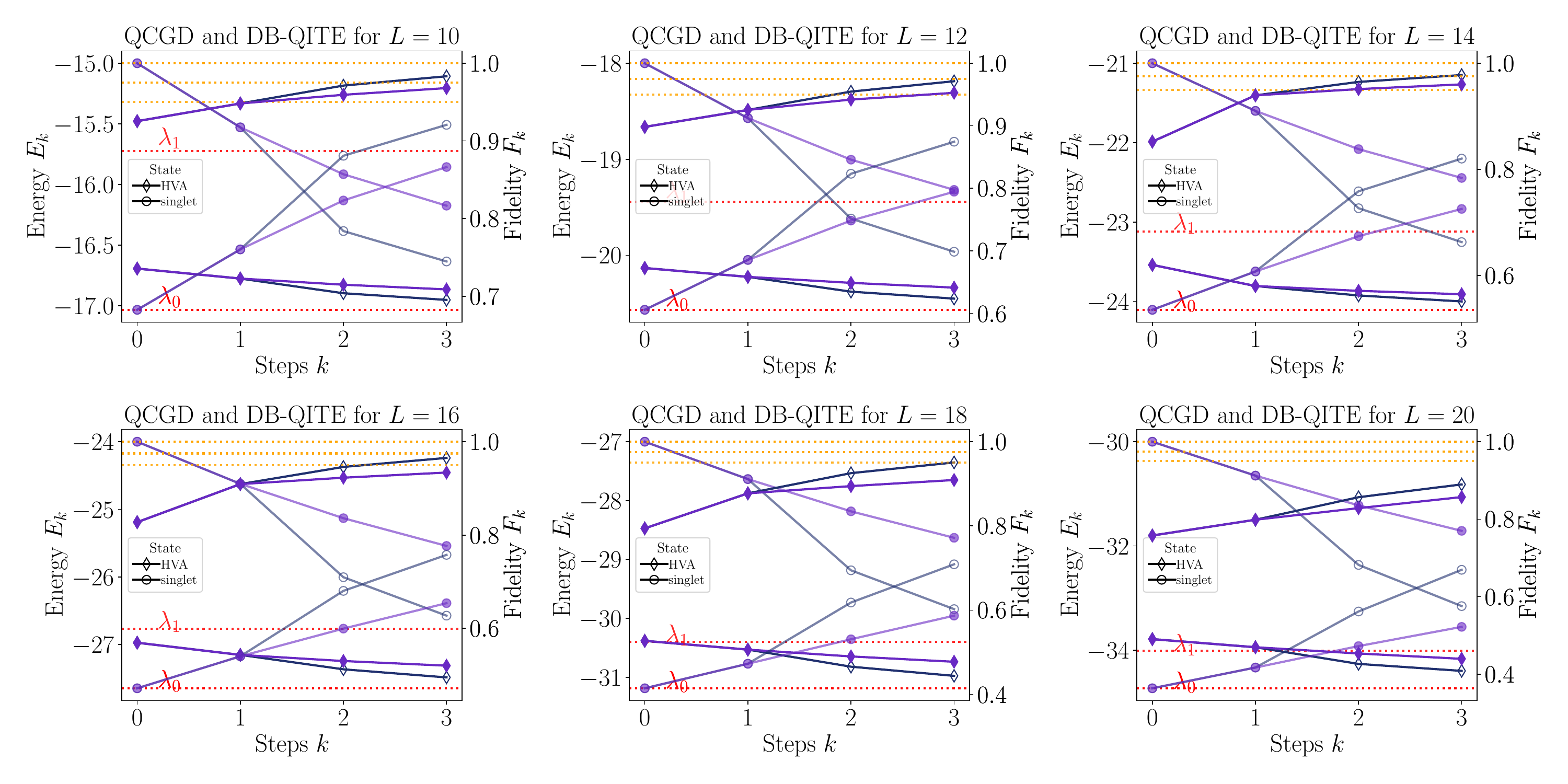}
    \caption{\textbf{Energy and ground-state fidelity for DB-QITE and QCGD over three updates of the \(L\)-qubit Heisenberg model.} Purple filled markers and navy blue hollow markers distinguish DB-QITE and QCGD respectively, while diamonds and circles differentiate between HVA and singlet initialisation. Orange dotted lines mark fidelities 1, 0.975, and 0.95; red dotted lines mark the ground-state and first excited state energies \(\lambda_0\) and \(\lambda_1\).}
    \label{fig: energies and fidelities across L}
\end{figure*}

\begin{table}[ht]
    \centering
    \caption{\textbf{Percentage difference between the excess energy reduction of QCGD and DB-QITE}. We report $\Delta R_k$ after the second and third updates for both initialisations. All values shown here are given in percent.}
    \label{tab:comparison}
\begin{tabular}{cc cccccc}
\toprule
$\ket{\psi_0}$ & $\Delta R_k$
& \multicolumn{6}{c}{System size \(L\)} \\
\cmidrule(lr){3-8}
& & \(10\) & \(12\) & \(14\) & \(16\) & \(18\) & \(20\) \\
\midrule
\multirow{2}{*}{Singlet}
& \(\Delta R_2\) & \(23.0\) & \(23.9\) & \(24.0\) & \(23.9\) & \(23.9\) & \(24.1\) \\
& \(\Delta R_3\) & \(22.6\) & \(25.1\) & \(26.0\) & \(28.3\) & \(28.9\) & \(30.5\) \\
\addlinespace
\multirow{2}{*}{HVA}
& \(\Delta R_2\) & \(20.7\) & \(21.0\) & \(10.3\) & \(18.2\) & \(21.6\) & \(21.1\) \\
& \(\Delta R_3\) & \(25.4\) & \(26.4\) & \(16.0\) & \(25.8\) & \(29.7\) & \(24.5\) \\
\bottomrule
\end{tabular}
\end{table}
We next examine the ground-state fidelity
\begin{equation}
    F_k^\mathcal{A} = \left|\braket{\lambda_0|\psi_k^\mathcal{A}}\right|^2.
\end{equation}
Fig.~\ref{fig: energies and fidelities across L} shows the energy and fidelity for all simulated sizes. While both algorithms improve the fidelity systematically, QCGD achieves a larger final value for every tested \(L\) and both initialisations.

The performance gap is most pronounced for the lower-quality singlet initialisation. At \(L=20\), QCGD raises the singlet fidelity to 0.669 compared to 0.522 for DB-QITE, giving an absolute increase of 0.147. For the HVA initial state, the corresponding values are 0.890 and 0.857, an increase of 0.033. Table~\ref{table: final fidelities} summarises the complete data. The final fidelity generally decreases with $L$ for both methods, with a small variation for the HVA initialization for \(L=14\). In comparison, the normalised excess energy advantage from Table~\ref{tab:comparison} is more stable. Therefore, energy-based criteria may be a more informative diagnostic for state preparation when a useful approximation only requires a modest overlap with the exact ground-state.

\begin{table}[t]
    \centering
    \caption{\textbf{Final ground-state fidelities \(F_3^\mathcal{A} = \left|\braket{\lambda_0|\psi_3^\mathcal{A}}\right|^2\) achieved after three iterations of DB-QITE and QCGD for the $L$-qubit Heisenberg model.} Results are reported for both the singlet and HVA initialisations.}
    \label{table: final fidelities}
\begin{tabular}{c cc cc}
\toprule
$L$ & \multicolumn{2}{c}{DB-QITE} & \multicolumn{2}{c}{QCGD} \\
\cmidrule(lr){2-3}\cmidrule(lr){4-5}
 & HVA & Singlet & HVA & Singlet \\
\midrule
10 & 0.968 & 0.866 & 0.983 & 0.921 \\
12 & 0.953 & 0.795 & 0.971 & 0.875 \\
14 & 0.960 & 0.725 & 0.978 & 0.820 \\
16 & 0.934 & 0.654 & 0.965 & 0.758 \\
18 & 0.909 & 0.587 & 0.951 & 0.709 \\
20 & 0.857 & 0.522 & 0.890 & 0.669 \\
\bottomrule
\end{tabular}
\end{table}
    
\subsubsection{Compiled-circuit resources}
\label{circuit resources}
\begin{figure*}
    \centering
    \includegraphics[scale=0.35]{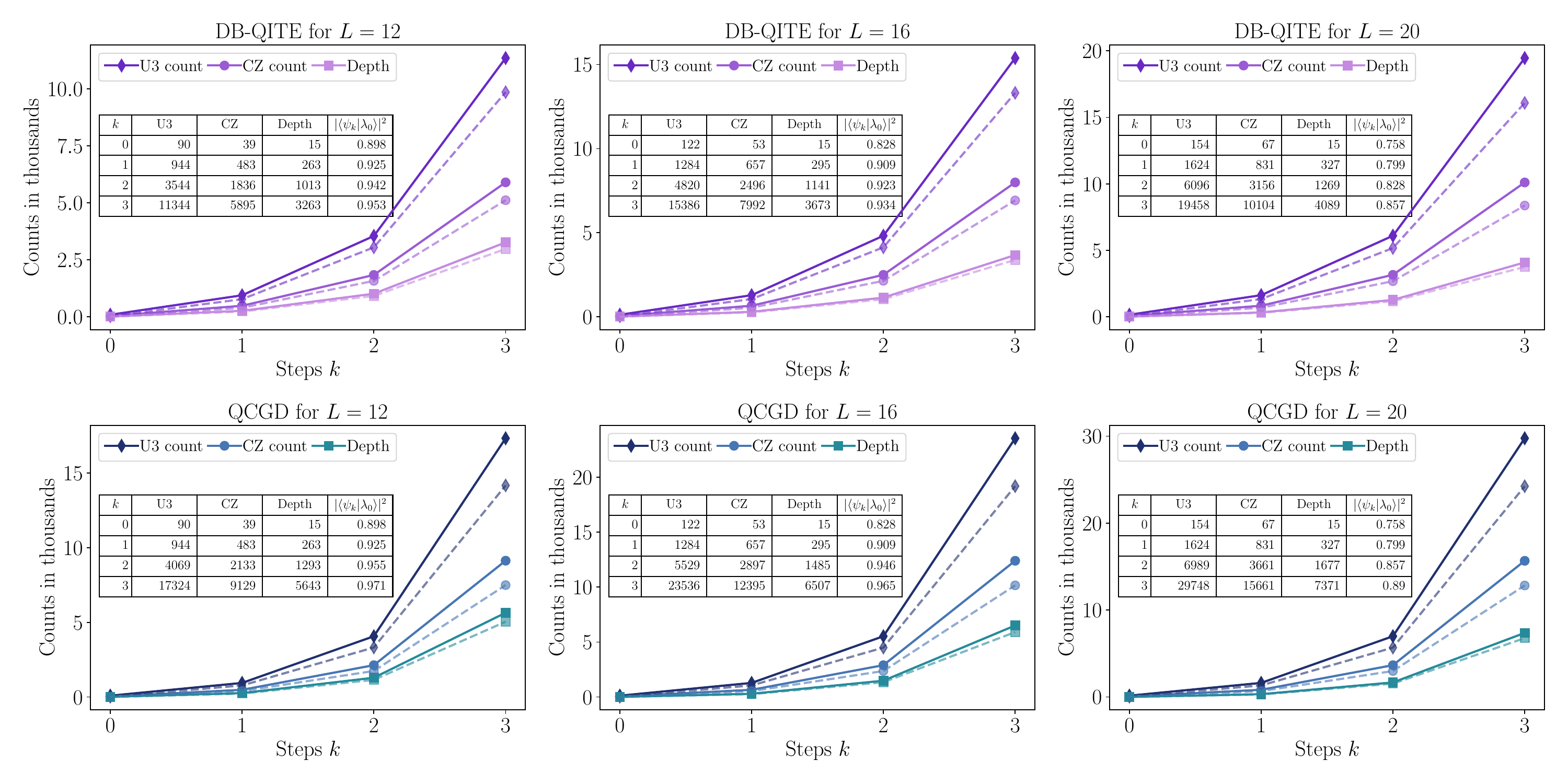}
    \caption{\textbf{Compiled gate counts, circuit depth, and ground-state fidelity for DB-QITE (purple palette) and QCGD (blue palette) at $L=12,16,20$.} The columns correspond to increasing system size. Solid curves show HVA initialisation, while dashed curves show singlet initialisation. Each inset reports the compiled $U_3$ and \(\mathrm{CZ}\) gate counts, circuit depth, and HVA fidelity at the corresponding update. At fixed $k$, QCGD uses a larger circuit, whereas the matched-fidelity comparison shows that two QCGD updates match or exceed the HVA fidelity reached after three DB-QITE updates for these representative system sizes; the corresponding resource ratios are given in Table~\ref{tab:matched-fidelity-resources}.}
    \label{fig: circuit data}
\end{figure*}

We now compare the elementary gates required to implement DB-QITE and QCGD. To this end, we compile every iteration into the same gate set and record the numbers of single-qubit \(U3\) gates and \(\mathrm{CZ}\) gates, together with the circuit depth. In Fig.~\ref{fig: circuit data} we show the resulting growth for DB-QITE and QCGD. At the same iteration count, QCGD is more expensive than DB-QITE due to the additional state reflections. For the HVA initialisation and \(L = 12,16,20\), the QCGD circuit at \(k=2\) uses approximately \(16\%\) more \(\mathrm{CZ}\) gates and between \(28\%\) and \(32\%\) more depth than DB-QITE. At $k=3$, the overhead increases substantially. Thus, as established above, improved convergence per iteration does not, by itself, imply a practical advantage.

The relevant conclusion emerges when comparing both algorithms at matched fidelity. For \(L=12\), two QCGD iterations achieve \(F_2 = 0.955\), which slightly exceeds the final DB-QITE value \(F_3 = 0.953\). Similarly, for \(L=16\) we have \(0.946\) compared to 0.934, and for \(L=20\) QCGD obtains a fidelity of 0.857 one iteration earier. Thus, QCGD is able to match or improve the fidelity with one fewer step while requiring only \(35.9\%\) of the \(U3\) gates, \(36.2\%\) of the \(\mathrm{CZ}\) gates, and \(39.6\%\) to \(41.0\%\) of the depth of the corresponding three-step DB-QITE circuits. These results, compiled in Table~\ref{tab:matched-fidelity-resources}, demonstrate that the higher cost of an individual QCGD update can be offset by avoiding one level of recursive circuit growth.
\subsubsection{Numerical limitations}
\label{limitations}

\begin{figure*}
    \centering
    \includegraphics[width=1\linewidth]{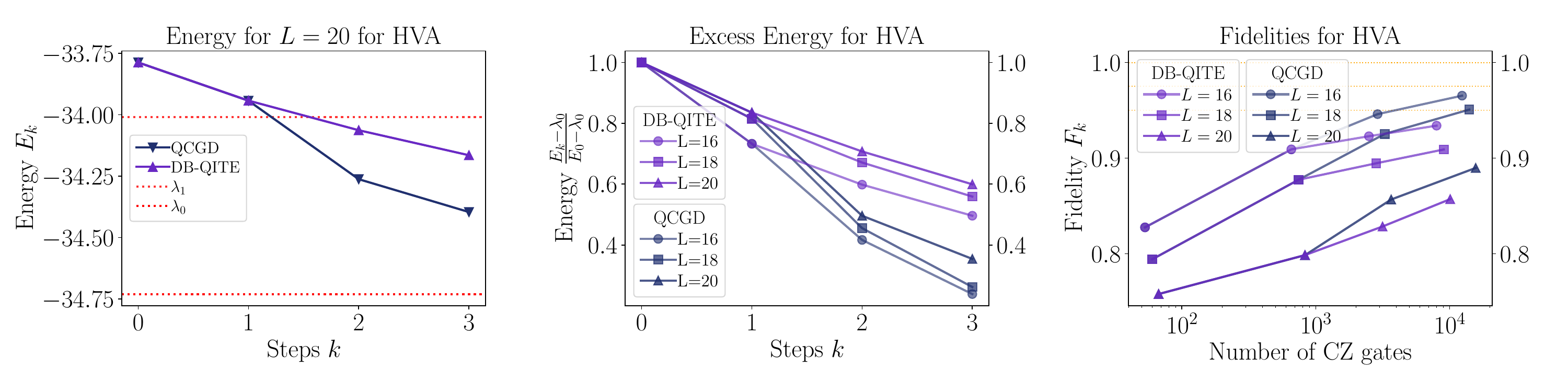}
     \includegraphics[width=1\linewidth]{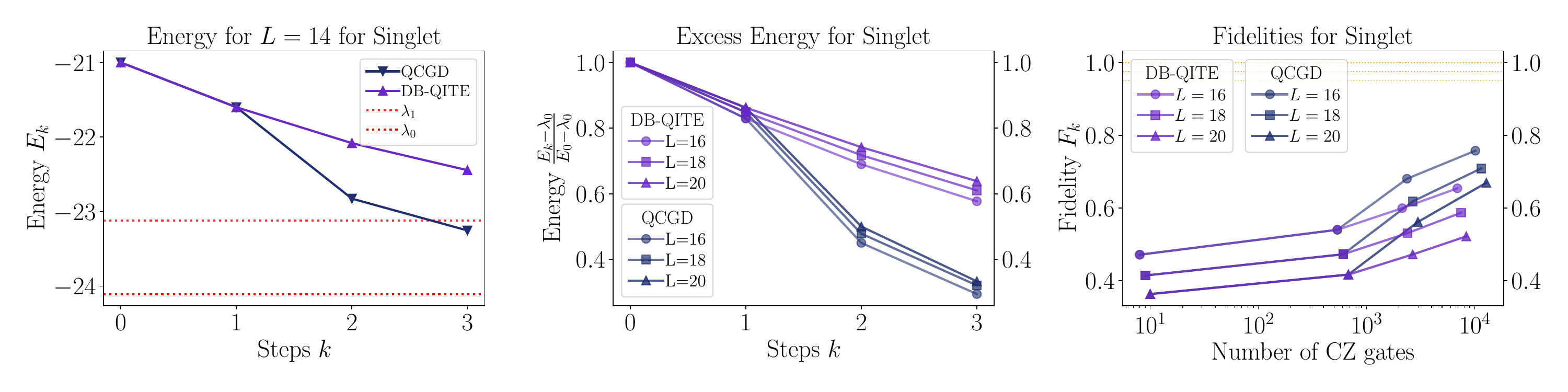}
    \caption{\textbf{Numerical benchmarks for the one-dimensional Heisenberg model.} The upper and lower rows correspond to the HVA and singlet initialisations, respectively. The left panels compare the representative DB-QITE and QCGD energy trajectories. The middle panels show the normalised excess energy for larger system sizes and exemplify that each algorithm performs consistently as \(L\) changes. The right panels report the ground-state fidelity against the number of compiled \(\mathrm{CZ}\) gates. QCGD  reliably removes a larger fraction of the excess energy over the tested trajectories, while the resource-based panels expose the cost of its more involved update.}
    \label{fig: representative data}
\end{figure*}

The presented benchmarks in Tables \ref{tab:comparison}, \ref{table: final fidelities}, and \ref{tab:matched-fidelity-resources} isolate the algorithmic and compiled-circuit results of the QCGD update, but certain numerical limitations remain unexplored. For instance, the two-dimensional parameter search used by QCGD requires more candidate circuits than the simpler DB-QITE line search. Moreover, as shown in Fig.~\ref{fig: heatmaps}, QCGD can benefit from an extended parameter domain, meaning that future work is needed to determine the optimal parameter range for a given system size \(L\). 

These results are restricted to the open Heisenberg chain, two physically motivated initialisations, and three recursive steps. Thus, they show that the conjugate direction can provide a consistent advantage under these conditions, but they do not guarantee uniform improvement for arbitrary Hamiltonians or further iterations. Thus, beyond these setting changes, future numerical work should test the robustness of the matched-fidelity resource advantage under finer, possibly extended, parameter grids and additional Hamiltonian families.

\begin{table}
    \centering
     \caption{\textbf{Matched-fidelity comparison between two QCGD iterations and three DB-QITE iterations for HVA initialisation.} The last three columns report the QCGD resource as a percentage of the corresponding DB-QITE resource.}
    \label{tab:matched-fidelity-resources}
    \begin{tabular}{c cc ccc}
        \toprule
        $L$
        & $F_2^{\mathrm{QCGD}}$
        & $F_3^{\mathrm{DB\text{-}QITE}}$
        & $\frac{N_{U3}^{\mathrm{QCGD}}}{N_{U3}^{\mathrm{DB\text{-}QITE}}}$
        & $\frac{N_{\mathrm{CZ}}^{\mathrm{QCGD}}}{N_{\mathrm{CZ}}^{\mathrm{DB\text{-}QITE}}}$
        & $\frac{D^{\mathrm{QCGD}}}{D^{\mathrm{DB\text{-}QITE}}}$ \\
        \midrule
        $12$ & $0.955$ & $0.953$ & $35.9\%$ & $36.2\%$ & $39.6\%$ \\
        $16$ & $0.946$ & $0.934$ & $35.9\%$ & $36.2\%$ & $40.4\%$ \\
        $20$ & $0.857$ & $0.857$ & $35.9\%$ & $36.2\%$ & $41.0\%$ \\
        \bottomrule
    \end{tabular}
\end{table}

\section{Discussion}
This work makes two main contributions. First, we extended the geometric formulation of QITE from Riemannian steepest descent~\cite{gluza_DB_QITE_2024} to a memory-augmented conjugate-gradient method. This construction yields QCGD, a quantum-compatible recursive update expressed in terms of Hamiltonian evolutions and reflections generated by previously prepared states. Second, our numerical and compiled-circuit benchmarks in Fig.~\ref{fig: representative data} and Fig.~\ref{fig: circuit data} demonstrate that this additional search direction can produce both faster convergence per iteration and a substantial resource advantage at matched fidelity.

For the Heisenberg model studied here, with \(L \in \{10,12,14,16,18,20\}\) qubits and two physically motivated initialisations, Fig.~\ref{fig: representative data} shows that QCGD removed a larger fraction of the initial excess energy after both the second and third updates for the tested trajectories. It also achieved a higher final ground-state fidelity than DB-QITE for every tested system size and initialisation. Because the state-preparation circuits grow recursively with the number of updates, the matched-fidelity comparison is more informative than a comparison at fixed iteration count, \(k\). At fixed \(k\), the QCGD state-preparation circuit is larger because it contains additional state reflections. Nevertheless, for the HVA trajectories at \(L=12,16,\) and \(20\), Table \ref{tab:matched-fidelity-resources} shows that two QCGD updates matched or exceeded the fidelity obtained after three DB-QITE updates. Relative to the corresponding three-step DB-QITE circuits, these two-step QCGD circuits used \(35.9\%\) of the \(U_3\) gates, \(36.2\%\) of the CZ gates, and \(39.6\%\)-\(41.0\%\) of the circuit depth. Therefore, in the tested regime, the improved search direction compensates for its higher fixed-step cost by avoiding one level of recursive circuit growth.

This numerical study aimed to determine whether a conjugate-gradient direction retains its optimization benefit, as illustrated in Fig.~\ref{fig: cgd vs gd}, after conversion to a quantum-compatible update, and whether that benefit offsets the resulting circuit overhead. The results throughout Section \ref{numerical results} establish both properties for the tested Heisenberg model and corresponding initialisation settings. Natural next steps include testing the robustness of the matched-fidelity resource advantage in broader settings, such as different simulation settings and potential algorithmic modifications, and deriving QCGD analogues of the cooling and ground-state-fidelity guarantees available for DB-QITE \cite{gluza_DB_QITE_2024}. Overall, the present results establish QCGD as a concrete demonstration that memory-based Riemannian optimization can translate into a substantial state-preparation circuit advantage for quantum GSP.

\vfill \eject

%%%%%%%%%%%%%%%%%%%%%%%%%%%%%%%%%%%%%%%%%%%%%%%%%%%%%%%%%%%%%%%%%%%%%%
%                                                                             
% REFERENCES
%                                                                             
%%%%%%%%%%%%%%%%%%%%%%%%%%%%%%%%%%%%%%%%%%%%%%%%%%%%%%%%%%%%%%%%%%%%%%
\clearpage
\input{qcgd.bbl}

% Single-column appendices
\clearpage
\onecolumngrid
\appendix
\subsection{Riemannian Gradient of the Energy}
\label{riemannian gradient computation}

This subsection derives the gradient introduced in Section \ref{geometric-formulation-for-ground-state-preparation} and verifies that its negative is a local energy-descent direction. For
\(
E(U)=\operatorname{Tr}(HU\rho_0U^\dagger)
\), consider the geodesic $U(t)=e^{tW}U_0$, where $W\in\mathfrak{u}(N)$, and define \(\psi_0=U_0\rho_0U_0^\dagger\). Proceeding from Eq.~\eqref{part1.eq.cost-change},
\begin{equation}
    \label{appendix eq cost-change}
    \Delta_W E(U(t))
    =\frac{E(e^{tW}U_0)-E(U_0)}{t},
\end{equation}
we expand the numerator to first order in $t$ to obtain
\begin{equation}
    \Delta_W E(U(t))
    =\operatorname{Tr}\!\left(H[W,\psi_0]\right)+O(t),
\end{equation}
where $[A,B]=AB-BA$. Hence,
\begin{equation}
    \left.\frac{d}{dt}E(U(t))\right|_{t=0}
    =\operatorname{Tr}\!\left(H[W,\psi_0]\right).
\end{equation}

Let us equip $T_{U_0}\mathcal{U}(N)$ with the Hilbert--Schmidt metric
\(
\langle A,B\rangle_{U_0}=\operatorname{Re}\operatorname{Tr}(A^\dagger B)
\).
Using cyclicity of the trace and expanding the commutator, we obtain
\begin{align}
    \left.\frac{d}{dt}E(U(t))\right|_{t=0}
    &=\operatorname{Tr}\!\left(W[\psi_0,H]\right) \\
    &=\operatorname{Tr}\!\left([\psi_0,H]WU_0U_0^\dagger\right) \\
    &=\operatorname{Tr}\!\left(\bigl([H,\psi_0]U_0\bigr)^\dagger WU_0\right) \\
    &=\left\langle[H,\psi_0]U_0,WU_0\right\rangle_{U_0}.
\end{align}
Therefore, \(\operatorname{grad}E(U_0)=[H,\psi_0]U_0\in T_{U_0}\mathcal{U}(N)\),
with $[H,\psi_0]\in\mathfrak{u}(N)$. Choosing $W=-[H,\psi_0]$ thus gives the steepest local decrease used by DB-QITE and QCGD.

\subsection{Higher-Order Commutator Formulae}

Here we present the corresponding higher-order DB-QITE and QCGD updates, which reduce the product-formula error from $O(\mu^{3/2})$ to $O(\mu^2)$, for completeness and for the resource comparison below.

\subsubsection{Higher-Order DB-QITE}
\label{dbqite hopf}

As shown in \cite{robbiatiDBQA2026}, we can write
\begin{align}
    e^{-\mu[H,\psi_k]}
    &=e^{i\sqrt{\mu}\phi H}
      e^{i\sqrt{\mu}\phi \psi_k}
      e^{-i\sqrt{\mu}H}
      e^{-i\sqrt{\mu}(1+\phi)\psi_k}
      e^{i\sqrt{\mu}(1-\phi)H}e^{i\sqrt{\mu} \psi_k}
      +O(\mu^2),
\end{align}
where  \(\phi=(\sqrt{5}-1)/2\). Since the right most factor contributes only a global phase when acting on \(U_k\), it can be omitted. Thus, implementing each $e^{i\beta\psi_k}$ as $U_ke^{i\beta\rho_0}U_k^\dagger$ yields
\begin{align}
\label{part2.eq.product-formula-2}
    U_{k+1}
    &=e^{i\sqrt{\mu}\phi H}
      U_ke^{i\sqrt{\mu} \phi\rho_0}U_k^\dagger
      e^{-i\sqrt{\mu}H}
      U_ke^{-i\sqrt{\mu}(1+\phi)\rho_0}U_k^\dagger
      e^{i\sqrt{\mu}(1-\phi)H}U_k.
\end{align}
\subsubsection{Higher-Order QCGD}
\label{qcgd hopf}
Applying the same group commutator approximation to the QCGD search direction yields
\begin{align}
    e^{-\mu[H,\phi_k]}
    &=e^{i\sqrt{\mu}\phi H}
      e^{i\sqrt{\mu}\phi \phi_k}
      e^{-i\sqrt{\mu}H}
      e^{-i\sqrt{\mu}(1+\phi)\phi_k}
      e^{i\sqrt{\mu}(1-\phi)H}e^{i\sqrt{\mu} \phi_k}
      +O(\mu^2),
\end{align}
where \(\phi_k = \sum_{j=0}^k c_j \psi_j\). Then, by using similar approximations as Eq.~\eqref{eq:reflection-sum-approx+} and Eq.~\eqref{eq:reflection-sum-approx-}, and using the modified coefficients \(\theta_j\) from Eq.~\eqref{eq:modified-coefficients} , we obtain, for $k\geq1$,
\begin{align}
\label{part3.eq.TGCI-2}
    U_{k+1}
    &=e^{i\phi\sqrt{\mu_k}H}
      e^{i\phi\sqrt{\mu_k}\psi_k}\cdots e^{i\phi \theta_0\psi_0}
      e^{-i\sqrt{\mu_k}H}
      e^{-i(1+\phi)\theta_0\psi_0}\cdots
      e^{-i(1+\phi)\sqrt{\mu_k}\psi_k}
      e^{i(1-\phi)\sqrt{\mu_k}H} e^{i \theta_0\psi_0}\cdots  e^{i\theta_{k-1}\psi_{k-1}}U_k,
\end{align}
where we have omitted a factor of \(e^{i \sqrt{\mu_k}\psi_k}\) since it contributes a global phase when acting on \(U_k\) as before. Here $U_1$ is produced by the higher-order DB-QITE update and \(\phi=(\sqrt{5}-1)/2\).

\subsection{Primitive-Query Complexity}
We count the number of primitive unitary operators; specifically, we denote by $N_{U_0}(k)$, $N_{\hat{H}}(k)$, and $N_{\hat{R}}(k)$ the number of operators corresponding to the initialisation unitary $U_0$, the Hamiltonian evolution operator $\hat{H}=e^{itH}$, and the phase rotation operator $\hat{R}=e^{it\rho_0}$, respectively.
\subsubsection{DB-QITE}
\label{dbqite query complexity}
We start with the query complexity of DB-QITE. We write the recursive relation defining the number of primitives \(N_A(k)\) of an operator \(A\) as 
\begin{equation}
\label{part2.eq.recursion-N}
N_A(k+1) = \alpha + \beta N_A(k), \, N_A(1) = \gamma,
\end{equation}
where \(\alpha\) is the number of times \(A\) appears in the update scheme, \(\beta\) is the number of appearances of \(U_k\), and \(\gamma\) is the number of gates \(A\) used in step \(k=1\). The solution to Eq.~\eqref{part2.eq.recursion-N} is
\begin{equation}
\label{part2.eq.recursion-solution}
    N_A(k)
    =-\frac{\alpha}{\beta-1}
     +\left(\gamma+\frac{\alpha}{\beta-1}\right)\beta^{k-1}.
\end{equation}

For the leading-order DB-QITE update, $U_k$ appears three times, so
\begin{alignat}{3}
    N_{U_0}(k+1)&=3N_{U_0}(k),       &\qquad N_{U_0}(1)&=3,\\
    N_{\widehat H}(k+1)&=2+3N_{\widehat H}(k), & N_{\widehat H}(1)&=2,\\
    N_{\widehat R}(k+1)&=1+3N_{\widehat R}(k), & N_{\widehat R}(1)&=1.
\end{alignat}
Equation~\eqref{part2.eq.recursion-solution} then gives
\begin{equation}
    N_{U_0}(k)=3^k,
    \qquad N_{\widehat H}(k)=3^k-1,
    \qquad N_{\widehat R}(k)=\frac{3^k-1}{2}.
\end{equation}

For the higher-order update in Eq.~\eqref{part2.eq.product-formula-2}, $U_k$ appears five times:
\begin{alignat}{3}
    N_{U_0}(k+1)&=5N_{U_0}(k),       &\qquad N_{U_0}(1)&=5,\\
    N_{\widehat H}(k+1)&=3+5N_{\widehat H}(k), & N_{\widehat H}(1)&=3,\\
    N_{\widehat R}(k+1)&=2+5N_{\widehat R}(k), & N_{\widehat R}(1)&=2,
\end{alignat}
and therefore
\begin{equation}
    N_{U_0}(k)=5^k,
    \qquad N_{\widehat H}(k)=\frac{3}{4}(5^k-1),
    \qquad N_{\widehat R}(k)=\frac{5^k-1}{2}.
\end{equation}
Thus, the leading- and higher-order DB-QITE circuits grow as $O(3^k)$ and $O(5^k)$, respectively.
\subsubsection{QCGD}
\label{qcgd query complexity}
Because the QCGD update contains earlier iterates, its primitive counts satisfy
\begin{align}
\label{part3.eq.gate-recursion}
    N_A(k+1)
    &=ak+b+c\sum_{j=1}^{k-1}N_A(j)+dN_A(k),
    \qquad N_A(1)=e,\quad N_A(2)=f,
\end{align}
where $ak+b$ counts explicit occurrences of $A$, while $c$ and $d$ count calls to earlier $U_j$ and to $U_k$, respectively. Eliminating the cumulative sum gives
\begin{equation}
    N_A(k+1)=a+(d+1)N_A(k)+(c-d)N_A(k-1).
\end{equation}

For the leading-order QCGD update,
\begin{alignat}{3}
    N_{U_0}(k+1)&=4N_{U_0}(k)+N_{U_0}(k-1),
        &\quad N_{U_0}(1)&=3,\;N_{U_0}(2)=13,\\
    N_{\widehat H}(k+1)&=4N_{\widehat H}(k)+N_{\widehat H}(k-1),
        & N_{\widehat H}(1)&=2,\;N_{\widehat H}(2)=8,\\
    N_{\widehat R}(k+1)&=2+4N_{\widehat R}(k)+N_{\widehat R}(k-1),
        & N_{\widehat R}(1)&=1,\;N_{\widehat R}(2)=6.
\end{alignat}
Writing $r_\pm=2\pm\sqrt{5}$, the solutions are
\begin{align}
    N_{U_0}(k)
    &=\frac{\left(\sqrt{5}+1\right)r_+^k + \left(\sqrt{5}-1\right)r_-^k}{2\sqrt{5}},\\
    N_{\widehat H}(k)
    &=\frac{r_+^k-r_-^k}{\sqrt{5}},\\
    N_{\widehat R}(k)
    &=-\frac{1}{2}
      +\frac{5-\sqrt{5}}{20}r_-^k
      +\frac{15+7\sqrt{5}}{20}r_+^{k-1}.
\end{align}

For the higher-order QCGD update,
\begin{alignat}{3}
    N_{U_0}(k+1)&=6N_{U_0}(k)+N_{U_0}(k-1),
        &\quad N_{U_0}(1)&=5,\;N_{U_0}(2)=31,\\
    N_{\widehat H}(k+1)&=6N_{\widehat H}(k)+N_{\widehat H}(k-1),
        & N_{\widehat H}(1)&=3,\;N_{\widehat H}(2)=18,\\
    N_{\widehat R}(k+1)&=3+6N_{\widehat R}(k)+N_{\widehat R}(k-1),
        & N_{\widehat R}(1)&=2,\;N_{\widehat R}(2)=15.
\end{alignat}
Writing $s_\pm=3\pm\sqrt{10}$, the solutions are
\begin{align}
    N_{U_0}(k)
    &=\frac{\left(5-\sqrt{10}\right)s_-^k+\left(5+\sqrt{10}\right)s_+^k}{10},\\
    N_{\widehat H}(k)
    &=\frac{-3 s_-^k + 3s_+^k}{2\sqrt{10}},\\
    N_{\widehat R}(k)
    &=-\frac{1}{2}
      +\frac{5-\sqrt{10}}{20}s_-^{k}
      +\frac{25+8\sqrt{10}}{20}s_+^{k-1}.
\end{align}
Hence, leading- and higher-order QCGD grow as
\(O((2+\sqrt{5})^k)\) and \(O((3+\sqrt{10})^k)\), respectively. These rates make explicit the trade-off studied in the main text, namely that QCGD uses a deeper recursive circuit per iteration in exchange for a potentially larger reduction in energy.

\end{document}